\documentclass[aps,prl,10pt,oneside,twocolumn,showpacs]{revtex4}

\usepackage{graphicx}
\usepackage{dcolumn}
\usepackage{epstopdf}
\usepackage{bm}
\usepackage{amssymb}
\usepackage{amsmath}
\usepackage{parskip}
\usepackage{multirow}
\usepackage{subfigure}
\usepackage[sort&compress]{natbib}
\newcommand{\degree}{\ensuremath{^\circ}}
\begin{document}

\title{Type I Superconductivity in ScGa$_3$ and LuGa$_3$ Single Crystals}
\author{E. Svanidze and E. Morosan}
\affiliation{Department of Physics and Astronomy, Rice University, Houston, Texas, 77005 USA}

\begin{abstract}
We present evidence of type I superconductivity in single crystals of ScGa$_3$ and LuGa$_3$, from magnetization, specific heat and resistivity measurements: low critical temperatures T$_c$ = 2.1 - 2.2 K, field-induced second-to-first order phase transition in the specific heat, critical fields less than 240 Oe and low Ginzburg-Landau coefficients $\kappa$ $\approx$ 0.23 and 0.30 for ScGa$_3$ and LuGa$_3$, respectively, are all traits of a type I superconducting ground state. These observations render ScGa$_3$ and LuGa$_3$ two of only several type I superconducting compounds, with most other superconductors being type II (compounds and alloys) or type I (elemental metals and metaloids).
\end{abstract}


\maketitle

\setlength{\parindent}{20pt}
\section{I Introduction}
Despite the large number of known conventional and unconventional superconductors (SCs), new findings still emerge even from simple, binary intermetallic systems. The majority of the metallic elements are superconducting with small values of the critical temperatures T$_c$.\cite{rob} It has been noted \cite{kak} that intermetallic compounds often have T$_c$ values higher than those of the constituent elements, as is the case in Nb$_3$Sn,\cite{mat3} V$_3$Si,\cite{mat2} ZrB$_2$ and NbB$_2$.\cite{gas} In this work, we present thermodynamic and transport measurements on single crystals of RGa$_3$ (R = Sc or Lu), formed with superconducting Ga with T$_c$ = 1.09 K,\cite{rob} and either non-superconducting Sc or superconducting Lu whose critical temperature is T$_c$ = 0.1 K.\cite{rob}

Past studies focused on the synthesis of polycrystalline samples of RGa$_3$, with reports on single crystals limited to de Haas van Alphen measurements.\cite{plu} Pluzhnikov \textsl{et. al.} characterized the geometry of the Fermi surface of three related intermetallic compounds, RGa$_3$ (R = Sc, Lu) and LuIn$_3$. Together with findings from band structure calculations \cite{afl} on the same systems, these reports suggested great similarities between the electronic properties of ScGa$_3$ and LuGa$_3$. Superconductivity below 2.3 K in LuGa$_3$ had already been mentioned,\cite{hav} but thermodynamic and transport properties measurements of both ScGa$_3$ and LuGa$_3$ have so far been limited to T $>$ 4.2 K.\cite{kleto, klet} The similarities in the electronic structures of ScGa$_3$ and LuGa$_3$ suggest that, if the superconductivity in the latter compound is confirmed, the former is likely to also display a superconducting ground state. In the current paper we show evidence that indeed both RGa$_3$ (R = Sc and Lu) are superconducting. The low critical temperatures T$_c$ around 2.2 K and small critical fields H$_{c}$ $<$ 240 Oe point to type I superconductivity in both these compounds. Additional supporting evidence for the type I superconductivity is provided by the field-dependent specific heat and low values of the Ginzburg-Landau (GL) coefficient $\kappa$ $\approx$ 0.23 and 0.3 for ScGa$_3$ and LuGa$_3$, respectively.

\begin{figure}
\centering
\includegraphics[width=\columnwidth]{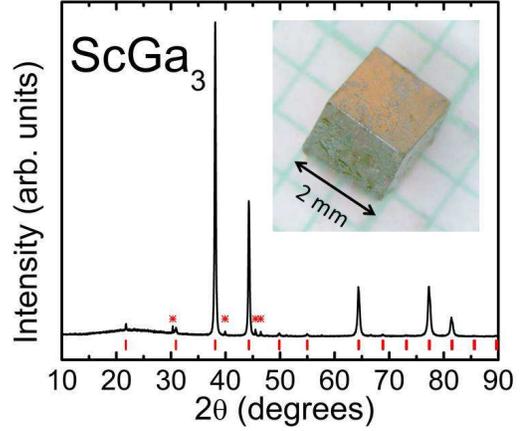}
\caption{powder X-ray pattern for ScGa$_3$ (black), with calculated peak positions (vertical red marks) for space group \textsl{Pm$\overline{3}$m} and lattice parameter a = 4.0919 \AA. Minute amounts of residual Ga flux are marked by asterisks. Inset: a picture of a single crystal of ScGa$_3$ prepared from a molten solution.}
\end{figure}

\begin{figure}
\centering
\includegraphics[width=0.8\columnwidth]{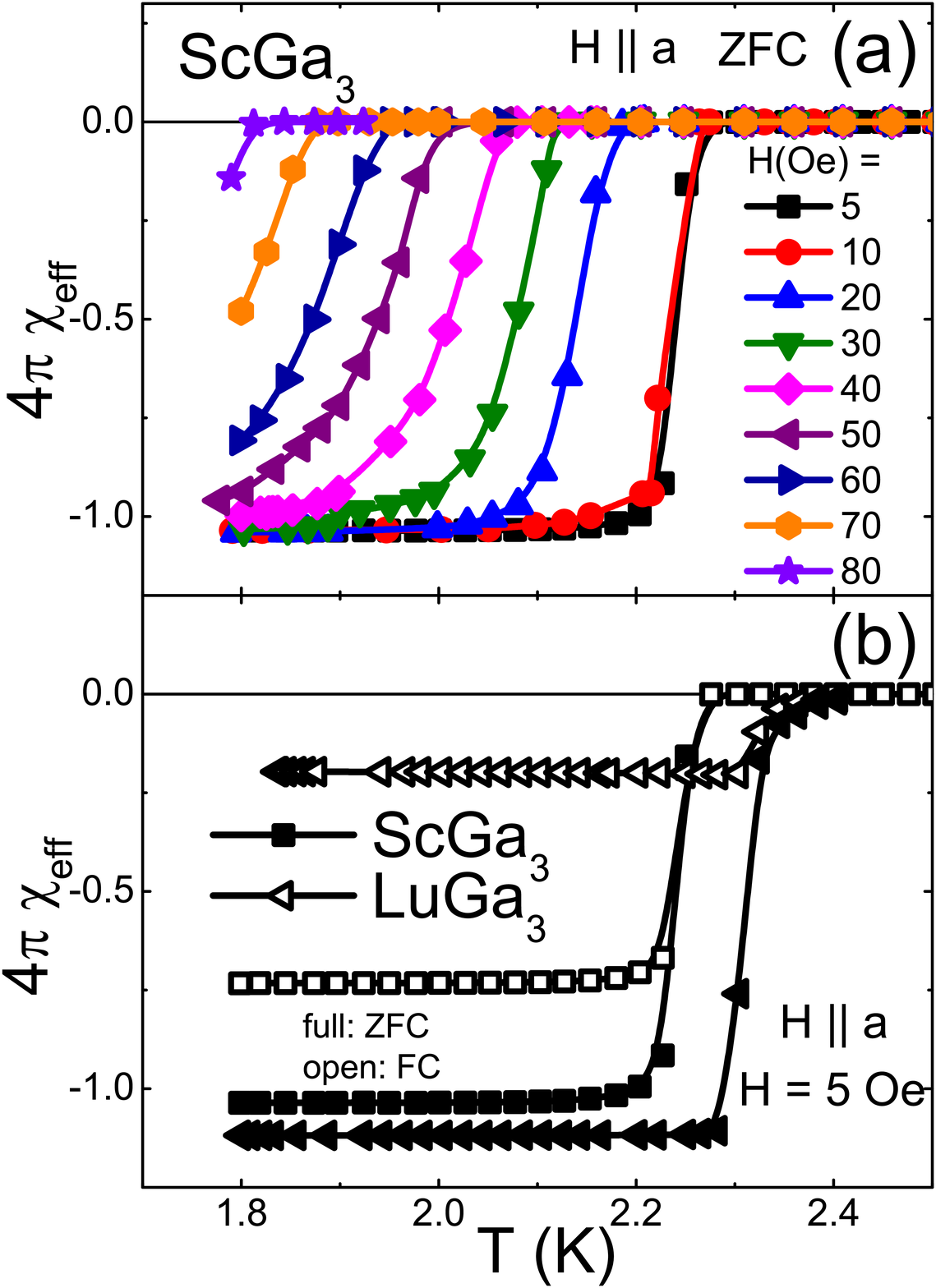}
\caption{(a) Zero-field cooled temperature-dependent susceptibility data, scaled by 4$\pi$ and corrected for demagnetizing effects 4$\pi$$\chi$$_{eff}$ = 4$\pi$$\chi$/(1- N$_d$$\chi$), for ScGa$_3$ in applied magnetic fields up to 80 Oe. (b) H = 5 Oe zero-field cooled (full symbols) and field-cooled (open symbols) scaled susceptibility 4$\pi$$\chi$$_{eff}$ data for ScGa$_3$ (squares) and LuGa$_3$ (triangles).}
\end{figure}

\section{II Experimental Methods}

The RGa$_3$ compounds (R = Sc, Dy-Tm, Lu) crystallize in the cubic \textsl{Pm$\overline{3}$m} space group, a structure suggested by Matthias \cite{mat} to be favorable for superconductivity. Single crystals of ScGa$_3$ and LuGa$_3$ were prepared using a self flux method by combining Sc or Lu (Hefa Rare Earth 99.999\%) with Ga (Alfa Aesar 99.9999\%). A R:Ga ratio of 1:9 was mixed in an alumina crucible, heated up to 930\degree C, then slowly cooled down to 760\degree C, followed by decanting of the residual flux in a centrifuge. Metallic cubic crystals with well-formed facets up to 2$\times$2$\times$2 \text{~mm}$^3$ in size were obtained. The crystals were then wrapped in Ta foil and annealed at 800\degree C for a week. Temperature- and field-dependent magnetization measurements with the magnetic field H parallel to the crystallographic axis H$||$a were performed in a Quantum Design (QD) Magnetic Property Measurement System, while specific heat data were collected in a QD Physical Property Measurement System (PPMS) using an adiabatic relaxation method. AC resistivity measurements from 0.4 K to 300 K were carried out using the standard four-probe method in the QD PPMS, with the current along the a axis \textsl{i} = 0.5 mA and f = 17.77 Hz. 

Powder X-ray diffraction data, shown in Fig. 1 for ScGa$_3$, were collected for both compounds in a Rigaku D/Max diffractometer using Cu K$\alpha$ radiation. The patterns for ScGa$_3$ and LuGa$_3$ were refined with the cubic space group \textsl{Pm}$\overline{3}$\textsl{m}, with lattice parameters a = 4.09 $\AA$ and a = 4.19 $\AA$, respectively. A picture of a ScGa$_3$ crystal is also shown in the inset in Fig. 1. Traces of residual Ga flux are apparent in the powder pattern, and are marked with asterisks in Fig. 1. Additional single crystal X-ray diffraction measurements confirmed the crystals structure, stoichiometry and purity of the ScGa$_3$ crystals.

\section{III Results and Discussion}
As-measured susceptibility data $\chi$ = M/H for RGa$_3$ in various applied magnetic fields H was scaled by 4$\pi$ and corrected for demagnetizing effects 4$\pi$$\chi$$_{eff}$ = 4$\pi$$\chi$/(1 - N$_d$$\chi$) as shown in Fig. 2. The demagnetizing factor, N$_d$ $\approx$ 1/3,\cite{aha, osb} is associated with the cubic geometry of the crystals. As anticipated from their electronic properties,\cite{plu} both R = Sc (Fig. 2a) and Lu (Fig. 2b) compounds display similar superconducting ground states below 2.2 - 2.3 K. Increasing magnetic field suppresses the transition for ScGa$_3$ (Fig. 2a), such that T$_c$ becomes smaller than 1.8 K for H $\approx$ 80 Oe. Fig. 2b illustrates the similarity between the H = 5 Oe M(T) data for ScGa$_3$ (squares) and LuGa$_3$ (triangles), for both zero-field cooled (full) and field-cooled (open) data. The critical field H$_{c}$ for each compound can also be estimated from the M(H) data, shown in Fig. 3. Taking the demagnetization effect into consideration, a more accurate estimate of the field H is H$_{eff}$ = H - N$_d$M, where, as before, for a cube and H$||$a, N$_d$ $\approx$ 1/3. The resulting M(H$_{eff}$) isotherms are displayed in Fig. 3 (full symbols, bottom axes) along with as-measured M(H) for T = 1.8 K (open symbols, top axes). The critical field values H$_{c}$, corresponding to the entrance to the normal state (M = 0), are not changed when demagnetizing effects are taken into account for H$||$a. The critical fields are remarkably low, H$_{c}$ reaching only about 90 Oe at 1.8 K, the lowest temperature available for the magnetization measurements. Moreover, as will be shown below, the critical fields for both compounds remain small down to 0.4 K. This observation, along with the small critical temperatures and the shape of the M(H) isotherms, indicate type I superconductivity in both ScGa$_3$ and LuGa$_3$. While most elemental SCs are type I, this is a rare occurrence in superconducting compounds, making ScGa$_3$ and LuGa$_3$ two of only a few such known systems.\cite{ana, yon, tsi, wak, zha} It is therefore imperious to fully characterize the superconducting state in the RGa$_3$ SCs. Specific heat and resistivity measurements allow us to extend the findings from magnetization data down to lower temperatures.


\begin{figure}[t]
\centering
\includegraphics[width=0.8\columnwidth]{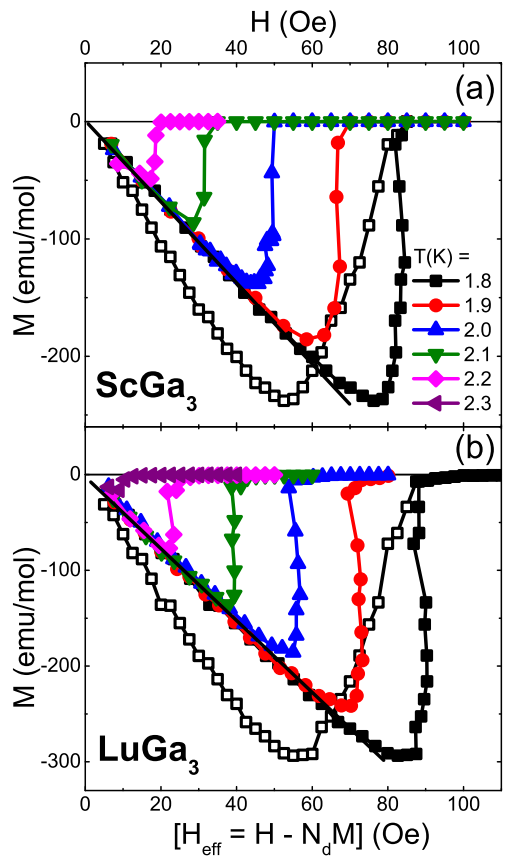}
\caption{(a) ScGa$_3$ and (b) LuGa$_3$ M(H$_{eff}$) for temperatures between 1.8 K and 2.3 K, where H$_{eff}$ = H - N$_d$M and N$_d$ is the demagnetizing factor for H$||$a. Open squares: M(H) isotherms for T = 1.8 K, where H is the applied (external) magnetic field.}
\end{figure}

Field-dependent specific heat measurements for ScGa$_3$ and LuGa$_3$ were carried out in fields up to 240 Oe, as shown in Fig. 4. As expected, a sharp peak is observed for field values H $<$ 240 Oe, from which the critical temperature T$_c$ can be determined as the point halfway between the peak and the normal state specific heat signal. Type I superconductivity in both compounds is confirmed by the increase of the jump in specific heat between zero and non-zero applied magnetic field H, indicating second-to-first order phase transition. T$_c$ for ScGa$_3$ and LuGa$_3$ is suppressed from 2.1 K (open squares, Fig. 4a) and 2.0 K (open squares, Fig. 4b), respectively, at H = 0 to below 0.4 K at H = 240 Oe (solid line, Fig. 4c and d). The normal state electronic specific heat coefficient $\gamma$$_n$ and phonon specific heat coefficient $\beta$ were estimated from the linear fit of the normal state (H = 240 Oe) specific heat below 8 K, plotted as C$_p$/T \textsl{vs.} T$^2$ (not shown). Very similar $\gamma$$_n$ values, 7.11 mJ mol$^{-1}$ K$^{-2}$ and 8.46 mJ mol$^{-1}$ K$^{-2}$, were obtained for ScGa$_3$ and LuGa$_3$, respectively. The experimental $\gamma$$_n$ values are larger than those estimated ($\gamma$$_{n,PPPW}$ = 2.4 mJ/mol K$^2$ for ScGa$_3$, and 1.2 mJ/mol K$^2$ for LuGa$_3$) from existing band structure calculations based on the Pseudo-Potential Plane Wave approximation (PPPW). \cite{afl} However, a more accurate estimate of $\gamma$$_n$ results from the Full Potential Linear Augmented Plane Wave method (FPLAPW),\cite{Jiakui} which gives $\gamma$$_{n,FPLAPW}$ = 7.1 mJ/mol K$^2$ for ScGa$_3$, identical with the experimental value of 7.11 mJ/mol K$^2$. The superconducting electronic specific heat coefficient $\gamma$$_s$ can also be determined from $\gamma$$_n$ and the residual electronic specific heat coefficient $\gamma$$_{res}$. The latter coefficient, $\gamma$$_{res}$, estimated from C$_e$/T at T = 0.4 K and H = 0 (Fig. 4c and d), is much smaller than $\gamma$$_n$ for both compounds. This results in $\gamma$$_s$ = $\gamma$$_n$ - $\gamma$$_{res}$ $\approx$ $\gamma$$_n$ for both ScGa$_3$ and LuGa$_3$. The entropy-conservation construct shown in Fig. 4c and d for ScGa$_3$ and LuGa$_3$, respectively, yields the same value for the jump in the electronic specific heat C$_e$ at T$_c$, $\Delta$C$_e$/$\gamma$$_n$T$_c$ $\approx$ 1.44, consistent with BCS-type superconductivity.\cite{bar} One more similarity between the two compounds is the minimum excitation energy $\Delta$(0): from the low-temperature fit of the electronic specific heat C$_e$ $\propto$ e $^{-\Delta/k_BT}$ (dashed lines in Fig. 4c and d), $\Delta$(0) is estimated to be 0.18 meV for ScGa$_3$ and 0.17 meV for LuGa$_3$. The Debye temperature $\theta$$_D$ = (12$\pi$$^4$N$_A$rk$_B$/5$\beta$)$^{1/3}$, where r = 4 is the number of atoms per formula unit, can be determined using the phonon specific heat coefficient $\beta$ (Table 1), also estimated from the linear fit of C$_p$/T \textsl{vs.} T$^2$ (not shown). This yields $\theta$$_D$ = 660 K for ScGa$_3$ and 232 K for LuGa$_3$. Moreover, the electron-phonon coupling constant $\lambda$$_{el-ph}$, can be determined using McMillan's theory:\cite{mac}

\begin{center}
$\lambda_{el-ph} = \frac{1.04 + \mu^* ln(\theta_D/1.45 T_c)}{(1 - 0.62 \mu^*)ln(\theta_D/1.45 T_c) - 1.04}$
\end{center}

\noindent where $\mu$$^*$ represents the repulsive screened Coulomb potential and is usually between 0.1 and 0.15. Setting $\mu$$^*$ = 0.13, $\lambda_{el-ph}$ = 0.45 and 0.55 for ScGa$_3$ and LuGa$_3$, respectively, imply that both compounds are weakly coupled SCs. 

\begin{figure}[t]
\centering
\includegraphics[width=\columnwidth]{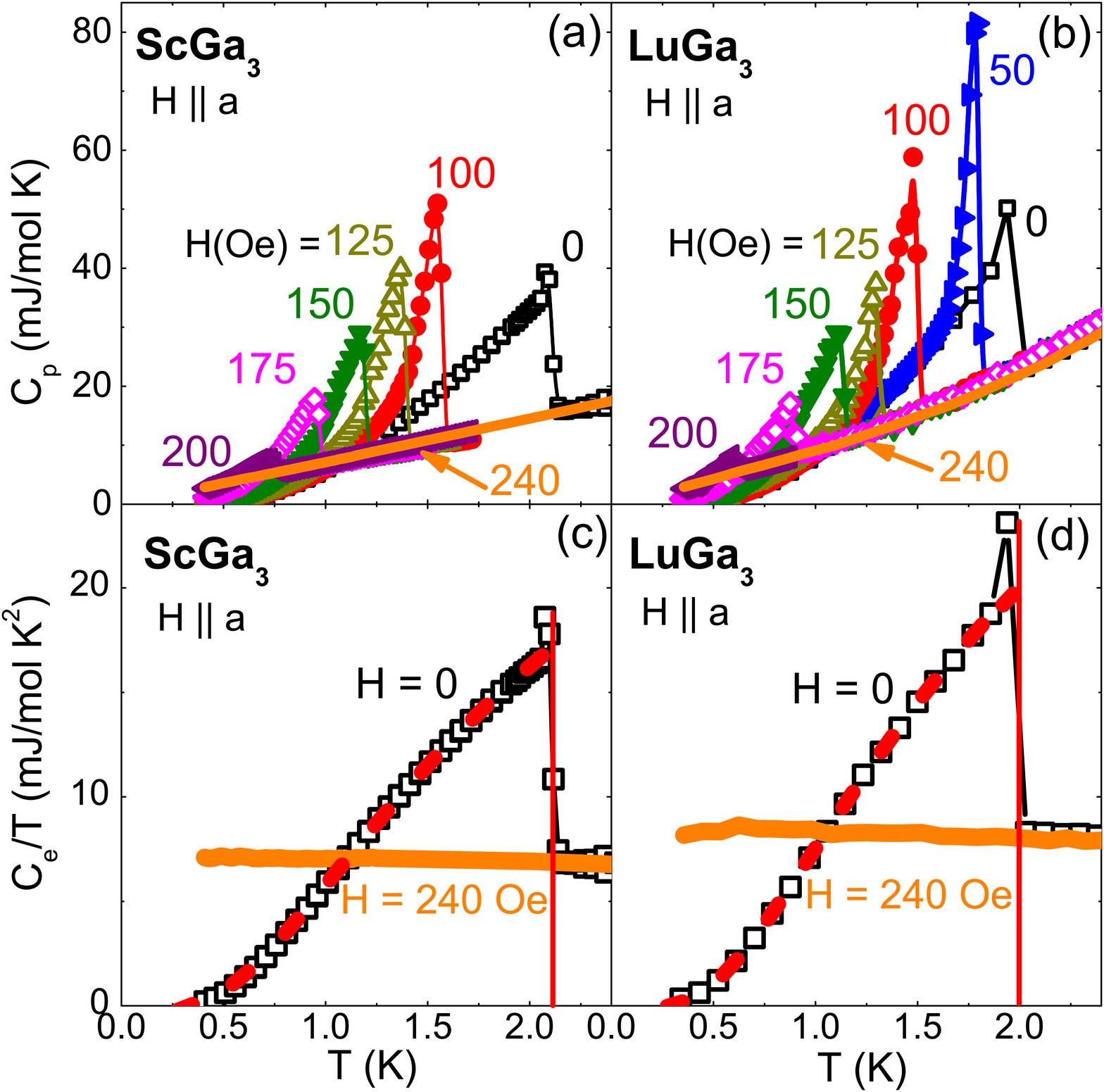}
\caption{Specific heat data for (a) ScGa$_3$ and (b) LuGa$_3$ in applied magnetic fields up 240 Oe. (c) and (d) Normal (H = 240 Oe) and superconducting (H = 0) electronic specific heat C$_e$, scaled by temperature T. The entropy conservation construct gives the ratio $\Delta$C$_{es}$/$\gamma$$_n$T$_c$ $\approx$ 1.44 for both ScGa$_3$ and LuGa$_3$, with the dashed line representing a fit of C$_e$/T to the expected BCS electronic specific heat.}
\end{figure}

From the specific heat data for both the superconducting (H = 0) and the normal (H = 240 Oe) states, an estimate of the thermodynamic critical field H$_c$ can be obtained using the free energy relation.\cite{tin} The thermodynamic critical field values H$_c$ = 209 $\pm$ 10 Oe for ScGa$_3$ and 226 $\pm$ 10 Oe for LuGa$_3$ are consistent with what has been observed in magnetization and specific heat data. The field- and temperature-dependent data can be summarized in the H - T phase diagram shown in Fig. 6 and discussed below. 

Previously reported resistivity measurements \cite{kleto, klet} on LuGa$_3$ were limited to temperatures above 4.2 K, while similar data had not been presented for ScGa$_3$. Fig. 5 displays the H = 0 resistivity data for ScGa$_3$ and LuGa$_3$ (full and open symbols, respectively). The superconducting transition (left inset) is around 2.2 - 2.3 K for both compounds. The apparently finite resistivity in the superconducting state is likely an artifact of the measurement: the overall resistivity values are very small for both compounds; below T$_c$, the contact resistance, albeit small, might alter the measured voltage, which is very close to the instrument resolution. Above the transition and below 80 K for ScGa$_3$ or 70 K for LuGa$_3$, $\rho$(T) exhibits Fermi liquid behavior, as illustrated by the $\Delta \rho$ $\propto$ AT$^2$ plot, with A = 3.4$\cdot$10$^{-4}$ and 6.1$\cdot$10$^{-4}$ $\mu$$\Omega$cmK$^{-2}$, respectively (right inset, Fig. 5). At higher temperatures a slight curvature of the resitivity is apparent. Fits to the Bloch-Gr\"{u}neisen-Mott (BGM) relation \cite{bid} (solid lines, Fig. 5)

\begin{center}
$\rho = \rho_0 + A (\frac{T}{\theta_D})^n\int_0^{\theta_D/T}\frac{x^ndx}{(e^x-1)(1-e^{-x})}-kT^3$
\end{center}

\noindent with n = 2 for ScGa$_3$ and n = 3 for LuGa$_3$ describe the data well up to room temperature, even higher than $\theta$$_D$/4. This points to significant s-d band scattering, while the different exponents n suggest underlying differences in the electron-phonon scattering in the two compounds. The fits shown in Fig. 5 were performed using the $\theta$$_D$ values determined from specific heat; the other BGM parameters were determined to be A = 38.5 and 28.6 $\mu$$\Omega$cm and k = 1.3$\cdot$10$^{-7}$ and 0.3$\cdot$10$^{-7}$ $\mu$$\Omega$cm/K$^3$, for ScGa$_3$ and LuGa$_3$, respectively. If the parameter $\theta$ is also released for the BGM fits, equally good fits for n = 2 and n = 3 are achieved for ScGa$_3$, for $\theta$$_R$ values between 320 and 460 K, significantly smaller than the Debye temperature $\theta$$_D$ = 660 K. For LuGa$_3$, the parameters remain nearly unchanged, with the best fit for n = 3 and $\theta$$_R$ = 230 K, virtually identical to $\theta$$_D$ = 232 K.

Based on the Sommerfield coefficient extracted from the specific heat data, it is possible to estimate the London penetration depth $\lambda$$_L$(0), the coherence length $\xi$(0) and the GL parameter $\kappa$(0) = $\lambda$$_L$(0)/$\xi$(0). Since both ScGa$_3$ and LuGa$_3$ have one formula unit per unit cell, the conduction electron density n, due to three electrons contributed by Sc and Lu, can be estimated as n = 3/V where V is the volume of the unit cell. It results that n = 4.39$\cdot$10$^{-2}$ \AA$^{-3}$ and n = 4.08$\cdot$10$^{-2}$ \AA$^{-3}$ for ScGa$_3$ and LuGa$_3$, respectively. If a spherical Fermi surface is assumed for both compounds, the Fermi wave vector k$_F$ can be roughly calculated as k$_F$ = (3n$\pi$$^2$)$^{1/3}$ = 1.09 \AA$^{-1}$ for ScGa$_3$ and 1.07 \AA$^{-1}$ for LuGa$_3$. The effective electron mass can then be determined as m* = $\hbar$$^2$k$^2_F$$\gamma$$_n$/$\pi$$^2$nk$^2_B$ = 3.03m$_0$ and 3.49m$_0$ for ScGa$_3$ and LuGa$_3$, respectively, where m$_0$ is the free electron mass. The London penetration depth is given as $\lambda$$_L$(0) = (m*/$\mu$$_0$ne$^2$)$^{1/2}$ = 59 nm for ScGa$_3$ and 63 nm for LuGa$_3$. The coherence length is then determined as $\xi$ = 0.18$\hbar$k$_F$/k$_B$T$_c$m* = 0.26 $\mu$m and 0.21 $\mu$m for ScGa$_3$ and LuGa$_3$, respectively. The GL parameter $\kappa$(0) = $\lambda$$_L$(0)/$\xi$(0) is thus 0.23 for ScGa$_3$ and 0.30 for LuGa$_3$. This indicates that both compounds are type I SCs, since $\kappa$ $<$ 1/$\sqrt{2}$. By comparison, MgB$_2$ is an example of a type II SC and its $\kappa$(0) is close to 26,\cite{fin} while $\kappa$(0) for LaRhSi$_3$, a reported type I superconducting compound, is close to 0.25.\cite{ana}

\begin{figure}
\centering
\includegraphics[width=\columnwidth]{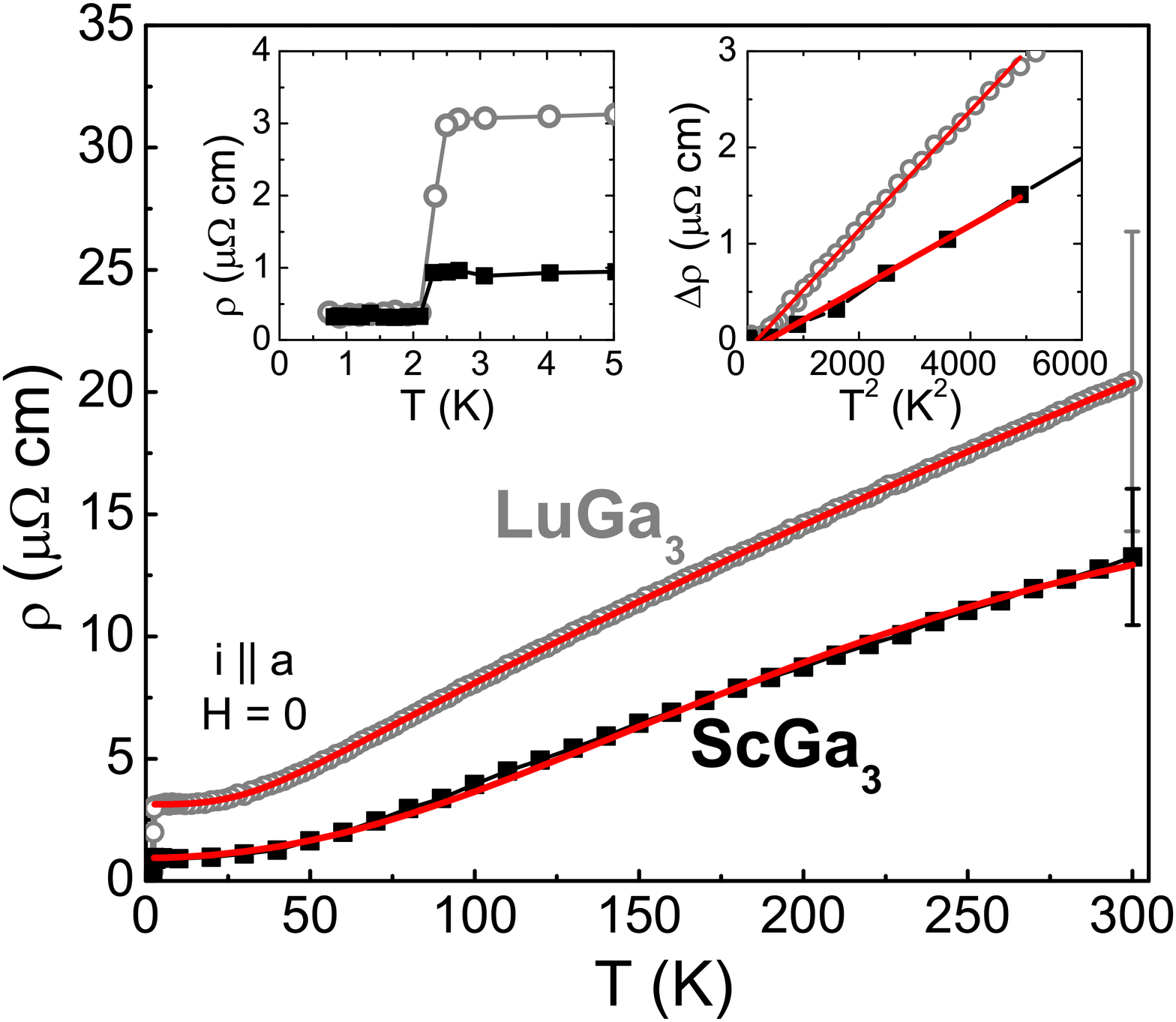}
\caption{H = 0 temperature-dependent resistivity for ScGa$_3$ (full black symbols) and LuGa$_3$ (open gray symbols), with Bloch-Gr\"{u}neisen-Mott fits (solid lines) for n = 2 (ScGa$_3$) and n = 3 (LuGa$_3$). Left inset: low-temperature $\rho$(T) around T$_c$. Right inset: $\Delta \rho$ = $\rho$ - $\rho$(0) \textsl{vs.} T$^2$, with solid lines representing linear fits up to 80 K for ScGa$_3$ and 70 K for LuGa$_3$.}
\end{figure}

\begin{figure}[t!]
\centering
\includegraphics[width=\columnwidth]{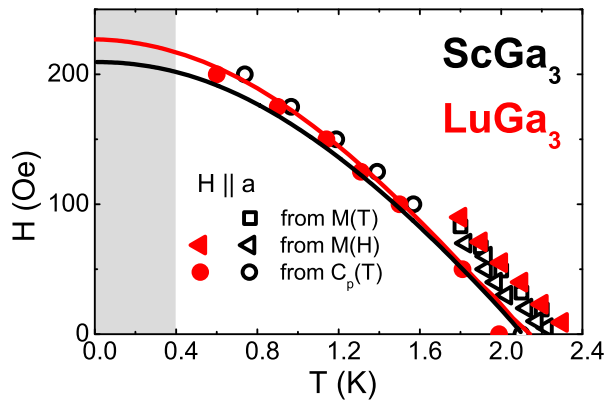}
\caption{H-T phase diagram for ScGa$_3$ (open black symbols) and LuGa$_3$ (full red symbols). The values of the critical fields H$_{c}$ are determined from M(T) data (squares), M(H) data (triangles) and C$_p$(T) data (circles).}
\end{figure}

\begin{center}
\begin{table*}
\caption{Summary of parameters describing ScGa$_3$ and LuGa$_3$ properties.}
  \begin{tabular}{ |c|c|c|c|c|c|c|c|c|c|c|c|c|c|}
    \hline
            &T$_c$         &H$_{c}$     &      $\gamma$$_n$        &$\beta$               &    A    &RRR &         $\Delta C_e(T_c)$   &$\lambda$$_{el-ph}$&m*             &$\lambda$ &$\xi$   & $\kappa$ \\ 
            & (K)          & (Oe)       &(mJmol$^{-1}$K$^{-2}$)&(mJmol$^{-1}$K$^{-4}$)&($\mu$$\Omega$cmK$^{-2}$)&     &$\overline{~~~\gamma_n T_c~~~}$&                   &(m$_0$)&(nm)      &($\mu$m)& \\ \hline
    ScGa$_3$&2.1 $\pm$ 0.2 &209 $\pm$ 10& 7.03 $\pm$ 0.08    & 0.027    &3.4$\cdot$10$^{-4}$                  &14.0&1.44                         & 0.45              &3.03           &59        &0.26    & 0.23 \\ \hline
    LuGa$_3$&2.2 $\pm$ 0.25&226 $\pm$ 10& 8.52 $\pm$ 0.06    & 0.621    &6.1$\cdot$10$^{-4}$                  &6.5 &  1.44                       & 0.55              &3.49           &63        &0.21    & 0.30  \\
    \hline

  \end{tabular}

\end{table*}
\end{center}

\section{IV Conclusions}
In summary, type I superconductivity in ScGa$_3$ and LuGa$_3$ is reported, with the parameters characteristic of the superconducting state shown in Table 1. The shape of the M(H) isotherms (Fig. 3), field-induced second-to-first order phase transition in specific heat (Fig. 4), low T$_c$, H$_c$ and $\kappa$ values (Table 1) suggest that ScGa$_3$ and LuGa$_3$ are both type I superconducting compounds. This is reflected also in the H - T phase diagram (Fig. 6), where the symbols represent experimental points from M(T) (squares), M(H) (triangles) and C$_P$ (circles). These data are in good agreement with the thermodynamic critical field H$_c$ temperature dependence (solid lines). As suggested by the electronic properties,\cite{plu} the superconducting parameters for the two compounds are very similar, as are their H - T phase diagrams. A careful analysis of the crystal structure on one hand, and the thermodynamic and transport properties of the type I superconducting compounds on the other hand, may offer valuable insights into the rare occurrence of type I superconductivity in binary or ternary systems. The relatively small electron-phonon coupling parameter $\lambda$$_{el-ph}$ indicates that both compounds are weakly-coupled BCS SCs. 


\section{V Aknowledgements}
This work was supported by NSF DMR 0847681. We thank A. Marcinkova, M. Beasley, A. Nevidomskyy, R. Prozorov and L. Zhao for useful discussions, J. Chan and G. McCandless for single crystal x-ray diffraction experiments and J. Wang for assistance with band structure calculations.


\bibliography{Sc-Ga_refs}

\begin{thebibliography}{24}
\expandafter\ifx\csname natexlab\endcsname\relax\def\natexlab#1{#1}\fi
\expandafter\ifx\csname bibnamefont\endcsname\relax
  \def\bibnamefont#1{#1}\fi
\expandafter\ifx\csname bibfnamefont\endcsname\relax
  \def\bibfnamefont#1{#1}\fi
\expandafter\ifx\csname citenamefont\endcsname\relax
  \def\citenamefont#1{#1}\fi
\expandafter\ifx\csname url\endcsname\relax
  \def\url#1{\texttt{#1}}\fi
\expandafter\ifx\csname urlprefix\endcsname\relax\def\urlprefix{URL }\fi
\providecommand{\bibinfo}[2]{#2}
\providecommand{\eprint}[2][]{\url{#2}}

\bibitem[{\citenamefont{Roberts}(1976)}]{rob}
\bibinfo{author}{\bibfnamefont{B.~W.} \bibnamefont{Roberts}},
  \bibinfo{journal}{Journal of Physical and Chemical Reference Data}
  \textbf{\bibinfo{volume}{5}}, \bibinfo{pages}{581} (\bibinfo{year}{1976}).

\bibitem[{\citenamefont{Kakani}(2006)}]{kak}
\bibinfo{author}{\bibfnamefont{S.~L.} \bibnamefont{Kakani}},
  \emph{\bibinfo{title}{Material Science}} (\bibinfo{year}{2006}).

\bibitem[{\citenamefont{Matthias et~al.}(1954)\citenamefont{Matthias, Geballe,
  Geller, and Corenzwit}}]{mat3}
\bibinfo{author}{\bibfnamefont{B.~T.} \bibnamefont{Matthias}},
  \bibinfo{author}{\bibfnamefont{T.~H.} \bibnamefont{Geballe}},
  \bibinfo{author}{\bibfnamefont{S.}~\bibnamefont{Geller}}, \bibnamefont{and}
  \bibinfo{author}{\bibfnamefont{E.}~\bibnamefont{Corenzwit}},
  \bibinfo{journal}{Physical Review} \textbf{\bibinfo{volume}{95}},
  \bibinfo{pages}{1435} (\bibinfo{year}{1954}).

\bibitem[{\citenamefont{Matthias}(1953)}]{mat2}
\bibinfo{author}{\bibfnamefont{B.~T.} \bibnamefont{Matthias}},
  \bibinfo{journal}{Physical Review} \textbf{\bibinfo{volume}{89}},
  \bibinfo{pages}{884} (\bibinfo{year}{1953}).

\bibitem[{\citenamefont{Gasparov et~al.}(2001)\citenamefont{Gasparov, Sidorov,
  Zver'kova, Khassanov, and Kulakov}}]{gas}
\bibinfo{author}{\bibfnamefont{V.~A.} \bibnamefont{Gasparov}},
  \bibinfo{author}{\bibfnamefont{N.~S.} \bibnamefont{Sidorov}},
  \bibinfo{author}{\bibfnamefont{I.~I.} \bibnamefont{Zver'kova}},
  \bibinfo{author}{\bibfnamefont{S.~S.} \bibnamefont{Khassanov}},
  \bibnamefont{and} \bibinfo{author}{\bibfnamefont{M.~P.}
  \bibnamefont{Kulakov}}, \bibinfo{journal}{Journal of Experimental and
  Theoretical Physics Letters} \textbf{\bibinfo{volume}{73}},
  \bibinfo{pages}{532} (\bibinfo{year}{2001}).

\bibitem[{\citenamefont{Pluzhnikov et~al.}(1995)\citenamefont{Pluzhnikov,
  Czopnik, and Svechkarev}}]{plu}
\bibinfo{author}{\bibfnamefont{V.~B.} \bibnamefont{Pluzhnikov}},
  \bibinfo{author}{\bibfnamefont{A.}~\bibnamefont{Czopnik}}, \bibnamefont{and}
  \bibinfo{author}{\bibfnamefont{I.~V.} \bibnamefont{Svechkarev}},
  \bibinfo{journal}{Physica B} \textbf{\bibinfo{volume}{212}},
  \bibinfo{pages}{375} (\bibinfo{year}{1995}).

\bibitem[{\citenamefont{http://www.aflowlib.org/index.html; S.~Curtarolo
  et~al.}(2003)\citenamefont{http://www.aflowlib.org/index.html; S.~Curtarolo,
  Morgan, Persson, Rodgers, and Ceder}}]{afl}
\bibinfo{author}{\bibnamefont{http://www.aflowlib.org/index.html;
  S.~Curtarolo}}, \bibinfo{author}{\bibfnamefont{D.}~\bibnamefont{Morgan}},
  \bibinfo{author}{\bibfnamefont{K.}~\bibnamefont{Persson}},
  \bibinfo{author}{\bibfnamefont{J.}~\bibnamefont{Rodgers}}, \bibnamefont{and}
  \bibinfo{author}{\bibfnamefont{G.}~\bibnamefont{Ceder}},
  \bibinfo{journal}{Physical Review Letters} \textbf{\bibinfo{volume}{91}},
  \bibinfo{pages}{135503} (\bibinfo{year}{2003}).

\bibitem[{\citenamefont{Havinga et~al.}(1970)\citenamefont{Havinga, Damsma, and
  van Maaren}}]{hav}
\bibinfo{author}{\bibfnamefont{E.~E.} \bibnamefont{Havinga}},
  \bibinfo{author}{\bibfnamefont{H.}~\bibnamefont{Damsma}}, \bibnamefont{and}
  \bibinfo{author}{\bibfnamefont{M.~H.} \bibnamefont{van Maaren}},
  \bibinfo{journal}{Journal of Physics and Chemistry of Solids}
  \textbf{\bibinfo{volume}{31}}, \bibinfo{pages}{2653} (\bibinfo{year}{1970}).

\bibitem[{\citenamefont{Kletowski}(1988)}]{kleto}
\bibinfo{author}{\bibfnamefont{Z.}~\bibnamefont{Kletowski}},
  \bibinfo{journal}{Physica Status Solidi} \textbf{\bibinfo{volume}{108}},
  \bibinfo{pages}{363} (\bibinfo{year}{1988}).

\bibitem[{\citenamefont{Kletowski et~al.}(1997)\citenamefont{Kletowski,
  Fabrowski, Slawinski, and Henkie}}]{klet}
\bibinfo{author}{\bibfnamefont{Z.}~\bibnamefont{Kletowski}},
  \bibinfo{author}{\bibfnamefont{R.}~\bibnamefont{Fabrowski}},
  \bibinfo{author}{\bibfnamefont{P.}~\bibnamefont{Slawinski}},
  \bibnamefont{and} \bibinfo{author}{\bibfnamefont{Z.}~\bibnamefont{Henkie}},
  \bibinfo{journal}{Journal of Magnetism and Magnetic Materials}
  \textbf{\bibinfo{volume}{166}}, \bibinfo{pages}{361} (\bibinfo{year}{1997}).

\bibitem[{\citenamefont{Matthias}(1954)}]{mat}
\bibinfo{author}{\bibfnamefont{B.~T.} \bibnamefont{Matthias}},
  \bibinfo{journal}{Physical Review} \textbf{\bibinfo{volume}{97}},
  \bibinfo{pages}{74} (\bibinfo{year}{1954}).

\bibitem[{\citenamefont{Aharoni}(1998)}]{aha}
\bibinfo{author}{\bibfnamefont{A.}~\bibnamefont{Aharoni}},
  \bibinfo{journal}{Journal of Applied Physics} \textbf{\bibinfo{volume}{83}},
  \bibinfo{pages}{3432} (\bibinfo{year}{1998}).

\bibitem[{\citenamefont{Osborn}(1945)}]{osb}
\bibinfo{author}{\bibfnamefont{J.~A.} \bibnamefont{Osborn}},
  \bibinfo{journal}{Physical Review} \textbf{\bibinfo{volume}{67}},
  \bibinfo{pages}{351} (\bibinfo{year}{1945}).

\bibitem[{\citenamefont{Anand et~al.}(2011)\citenamefont{Anand, Hillier,
  Adroja, Strydom, Michor, McEwen, and Rainford}}]{ana}
\bibinfo{author}{\bibfnamefont{V.~K.} \bibnamefont{Anand}},
  \bibinfo{author}{\bibfnamefont{A.~D.} \bibnamefont{Hillier}},
  \bibinfo{author}{\bibfnamefont{D.~T.} \bibnamefont{Adroja}},
  \bibinfo{author}{\bibfnamefont{A.~M.} \bibnamefont{Strydom}},
  \bibinfo{author}{\bibfnamefont{H.}~\bibnamefont{Michor}},
  \bibinfo{author}{\bibfnamefont{K.~A.} \bibnamefont{McEwen}},
  \bibnamefont{and} \bibinfo{author}{\bibfnamefont{B.~D.}
  \bibnamefont{Rainford}}, \bibinfo{journal}{Physical Review B}
  \textbf{\bibinfo{volume}{83}}, \bibinfo{pages}{064522}
  (\bibinfo{year}{2011}).

\bibitem[{\citenamefont{Yonezawa and Maeno}(2005)}]{yon}
\bibinfo{author}{\bibfnamefont{S.}~\bibnamefont{Yonezawa}} \bibnamefont{and}
  \bibinfo{author}{\bibfnamefont{Y.}~\bibnamefont{Maeno}},
  \bibinfo{journal}{Physical Review B} \textbf{\bibinfo{volume}{72}},
  \bibinfo{pages}{180504} (\bibinfo{year}{2005}).

\bibitem[{\citenamefont{Tsindlekht et~al.}(2006)\citenamefont{Tsindlekht,
  Leviev, Genkin, Felner, Paderno, and Filippov}}]{tsi}
\bibinfo{author}{\bibfnamefont{M.~I.} \bibnamefont{Tsindlekht}},
  \bibinfo{author}{\bibfnamefont{G.~I.} \bibnamefont{Leviev}},
  \bibinfo{author}{\bibfnamefont{V.~M.} \bibnamefont{Genkin}},
  \bibinfo{author}{\bibfnamefont{I.}~\bibnamefont{Felner}},
  \bibinfo{author}{\bibfnamefont{Y.~B.} \bibnamefont{Paderno}},
  \bibnamefont{and} \bibinfo{author}{\bibfnamefont{V.~B.}
  \bibnamefont{Filippov}}, \bibinfo{journal}{Physical Review B}
  \textbf{\bibinfo{volume}{73}}, \bibinfo{pages}{104507}
  (\bibinfo{year}{2006}).

\bibitem[{\citenamefont{Wakui et~al.}(2009)\citenamefont{Wakui, Akutagawa,
  Kase, Kawashima, Muranaka, Iwahori, Abe, and Akimitsu}}]{wak}
\bibinfo{author}{\bibfnamefont{K.}~\bibnamefont{Wakui}},
  \bibinfo{author}{\bibfnamefont{S.}~\bibnamefont{Akutagawa}},
  \bibinfo{author}{\bibfnamefont{N.}~\bibnamefont{Kase}},
  \bibinfo{author}{\bibfnamefont{K.}~\bibnamefont{Kawashima}},
  \bibinfo{author}{\bibfnamefont{T.}~\bibnamefont{Muranaka}},
  \bibinfo{author}{\bibfnamefont{Y.}~\bibnamefont{Iwahori}},
  \bibinfo{author}{\bibfnamefont{J.}~\bibnamefont{Abe}}, \bibnamefont{and}
  \bibinfo{author}{\bibfnamefont{J.}~\bibnamefont{Akimitsu}},
  \bibinfo{journal}{Journal of Physical Society of Japan}
  \textbf{\bibinfo{volume}{78}}, \bibinfo{pages}{034710}
  (\bibinfo{year}{2009}).

\bibitem[{\citenamefont{Zhao et~al.}(submitted, Physical Review
  B)\citenamefont{Zhao, Lausberg, Kim, Tanatar, Brando, Prozorov, and
  Morosan}}]{zha}
\bibinfo{author}{\bibfnamefont{L.~L.} \bibnamefont{Zhao}},
  \bibinfo{author}{\bibfnamefont{S.}~\bibnamefont{Lausberg}},
  \bibinfo{author}{\bibfnamefont{H.}~\bibnamefont{Kim}},
  \bibinfo{author}{\bibfnamefont{M.~A.} \bibnamefont{Tanatar}},
  \bibinfo{author}{\bibfnamefont{M.}~\bibnamefont{Brando}},
  \bibinfo{author}{\bibfnamefont{R.}~\bibnamefont{Prozorov}}, \bibnamefont{and}
  \bibinfo{author}{\bibfnamefont{E.}~\bibnamefont{Morosan}}
  (\bibinfo{year}{submitted, Physical Review B}).

\bibitem[{Jia(2012)}]{Jiakui}
\bibinfo{journal}{We performed band structure calculations for
  \text{Sc}\text{Ga}$_3$ using the full-potential linearized augmented
  plane-wave method (FPLAPW), as implemented in the \text{WIEN2K} code. This
  gives the \text{DOS}(\text{E}$_F$) $\approx$ 2 states/eV, from which it
  results that the electronic specific heat coefficient $\gamma$$_{FPLAPW}$ =
  7.1 mJ/mol K$^2$}  (\bibinfo{year}{2012}).

\bibitem[{\citenamefont{Bardeen et~al.}(1957)\citenamefont{Bardeen, Cooper, and
  Schrieffer}}]{bar}
\bibinfo{author}{\bibfnamefont{J.}~\bibnamefont{Bardeen}},
  \bibinfo{author}{\bibfnamefont{L.~N.} \bibnamefont{Cooper}},
  \bibnamefont{and} \bibinfo{author}{\bibfnamefont{J.~R.}
  \bibnamefont{Schrieffer}}, \bibinfo{journal}{Physical Review}
  \textbf{\bibinfo{volume}{108}}, \bibinfo{pages}{1175} (\bibinfo{year}{1957}).

\bibitem[{\citenamefont{McMillan}(1968)}]{mac}
\bibinfo{author}{\bibfnamefont{W.~L.} \bibnamefont{McMillan}},
  \bibinfo{journal}{Physical Review} \textbf{\bibinfo{volume}{167}},
  \bibinfo{pages}{331} (\bibinfo{year}{1968}).

\bibitem[{\citenamefont{Tinkham}(1996)}]{tin}
\bibinfo{author}{\bibfnamefont{M.}~\bibnamefont{Tinkham}},
  \emph{\bibinfo{title}{Introduction to Superconductivity}}
  (\bibinfo{year}{1996}).

\bibitem[{\citenamefont{Bid et~al.}(2006)\citenamefont{Bid, Bora, and
  Raychaudhuri}}]{bid}
\bibinfo{author}{\bibfnamefont{A.}~\bibnamefont{Bid}},
  \bibinfo{author}{\bibfnamefont{A.}~\bibnamefont{Bora}}, \bibnamefont{and}
  \bibinfo{author}{\bibfnamefont{A.~K.} \bibnamefont{Raychaudhuri}},
  \bibinfo{journal}{Physical Review B} \textbf{\bibinfo{volume}{74}},
  \bibinfo{pages}{035426} (\bibinfo{year}{2006}).

\bibitem[{\citenamefont{Finnemore et~al.}(2001)\citenamefont{Finnemore,
  Ostenson, Bud’ko, Lapertot, and Canfield}}]{fin}
\bibinfo{author}{\bibfnamefont{D.~K.} \bibnamefont{Finnemore}},
  \bibinfo{author}{\bibfnamefont{J.~E.} \bibnamefont{Ostenson}},
  \bibinfo{author}{\bibfnamefont{S.~L.} \bibnamefont{Bud’ko}},
  \bibinfo{author}{\bibfnamefont{G.}~\bibnamefont{Lapertot}}, \bibnamefont{and}
  \bibinfo{author}{\bibfnamefont{P.~C.} \bibnamefont{Canfield}},
  \bibinfo{journal}{Physical Review Letters} \textbf{\bibinfo{volume}{86}},
  \bibinfo{pages}{2420} (\bibinfo{year}{2001}).

\end{thebibliography}

\end{document}